**Main Manuscript for**

# High-resolution geostationary satellite observations of free tropospheric NO$_2$ over North America: implications for lightning emissions

Ruijun Dang[1*], Daniel J. Jacob[1], Huiqun Wang[2], Caroline R. Nowlan[2], Gonzalo Gonzalez Abad[2], Heesung Chong[2], Xiong Liu[2], Viral Shah[3,4], Laura H. Yang[1], Yujin J. Oak[1], Eloise A. Marais[5], Rebekah P. Horner[5], Andrew W. Rollins[6], James H. Crawford[7], Ke Li[8], and Hong Liao[8]

[1] John A. Paulson School of Engineering and Applied Sciences, Harvard University, Cambridge, MA 02138, USA.

[2] Atomic and Molecular Physics Division, Center for Astrophysics | Harvard & Smithsonian, Cambridge, MA 02138, USA.

[3] Global Modeling and Assimilation Office (GMAO), NASA Goddard Space Flight Center, Greenbelt, MD 20771, USA.

[4] GESTAR II, Morgan State University, Baltimore, MD 21251, USA.

[5] Department of Geography, University College London, London WC1E 6BT, UK.

[6] NOAA Chemical Sciences Laboratory, Boulder, CO 80305, USA

[7] NASA Langley Research Center, Hampton, VA 23666, USA.

[8] Jiangsu Key Laboratory of Atmospheric Environment Monitoring and Pollution Control, Collaborative Innovation Center of Atmospheric Environment and Equipment Technology, School of Environmental Science and Engineering, Nanjing University of Information Science and Technology, Nanjing 210044, China.

* Corresponding author: Ruijun Dang.

**Email:** rjdang@g.harvard.edu

**Author Contributions:** RD and DJJ conceptualized the research. RD conducted the research with contributions from DJJ, LHY, YJO, KL, and HL. HW, CRN, GGA, HC and XL provided guidance on the TEMPO satellite product. VS contributed to the interpretation of GEOS-CF results. EAM and RPH supplied the reprocessed TROPOMI cloud-sliced NO$_2$ product. AWR and JC offered guidance on aircraft observations. RD and DJJ wrote the paper, with input from all authors.

**Competing Interest Statement:** The authors declare that they have no conflict of interest.

**Classification:** Physical Sciences (major), Earth, Atmospheric, and Planetary Sciences (minor).

**This PDF file includes:**

Main Text
Figures 1 to 4

**Abstract (<250 words)**

Free tropospheric (FT) nitrogen dioxide ($NO_2$) plays a critical role in atmospheric oxidant chemistry as a source of tropospheric ozone and of the hydroxyl radical (OH). It also contributes significantly to satellite-observed tropospheric $NO_2$ columns, and must be subtracted when using these columns to quantify surface emissions of nitrogen oxide radicals ($NO_x \equiv NO + NO_2$). But large uncertainties remain in the sources and chemistry of FT $NO_2$ because observations are sparse. Here, we construct a new cloud-sliced FT $NO_2$ (700-300 hPa) product from the TEMPO geostationary satellite instrument over North America. This product provides higher data density and quality than previous products from low Earth orbit (LEO) instruments, with the first observation of the FT $NO_2$ diurnal cycle across seasons. Combined with coincident observations from the Geostationary Lightning Mapper (GLM), the TEMPO data demonstrate the dominance of lightning as a source of FT $NO_2$ in non-winter seasons. Comparison of TEMPO FT $NO_2$ data with the GEOS-CF atmospheric chemistry model shows overall consistent magnitudes, seasonality, and diurnal variation, with a midday minimum in non-winter seasons from photochemical loss. However, there are major discrepancies that we attribute to GEOS-CF's use of a standard cloud-top-height (CTH)-based scheme for the lightning $NO_x$ source. We find this scheme greatly underestimates offshore lighting flash density and misrepresents the diurnal cycle of lightning over land. Our FT $NO_2$ product provides a unique resource for improving the lightning $NO_x$ parameterization in atmospheric models and the ability to use $NO_2$ observations from space to quantify surface $NO_x$ emissions.

**Significance Statement (<120 words)**

Free tropospheric (FT) nitrogen dioxide ($NO_2$) drives global atmospheric oxidant chemistry, but its observations are sparse and its sources and chemistry are uncertain. Applying cloud-slicing to the TEMPO geostationary satellite instrument, we produce a unique high-quality FT $NO_2$ dataset over North America including the first measurements of the diurnal cycle. Using coincident geostationary lightning observations, we show that lightning dominates the FT $NO_2$ source in non-winter seasons and that atmospheric chemistry model parameterizations of that source have large errors. Correcting these errors is essential for models to properly describe global oxidant chemistry.

**Main Text**

**Introduction**

The free troposphere (FT), extending from the top of the planetary boundary layer (PBL) at about 2 km altitude to the tropopause, accounts for most of total atmospheric mass. It largely determines the abundance of tropospheric ozone and of the hydroxyl radical (OH), the main atmospheric oxidant, through chemistry catalyzed by nitrogen oxide radicals ($NO_x \equiv NO + NO_2$) (1-4). Sources of FT $NO_x$ include lightning, aviation, lofted wildfire plumes, frontal and deep convective lifting of surface $NO_x$ originating from fuel combustion and soils, and stratospheric downwelling. Lightning is thought to dominate the supply of FT $NO_x$ except in winter (5-7). However, there are large uncertainties in the magnitude and distribution of this lightning source (8-13). FT $NO_x$ chemistry is also not well understood, complicated by aerosol processes and organic nitrate reservoirs (6, 14-16). Addressing these uncertainties is critical for understanding tropospheric oxidant chemistry.

Progress in understanding $NO_x$ sources and chemistry in the FT has been limited by the sparsity of observations. In situ measurements are only available from aircraft with limited spatial and temporal coverage. Cloud slicing of $NO_2$ columns measured from satellites can provide more extensive coverage. It involves retrieving $NO_2$ columns over adjacent fully cloudy scenes with different cloud pressures. The retrieved $NO_2$ column is mainly restricted to the atmosphere above cloud top, so that a regression of $NO_2$ columns versus cloud pressure gives a measure of FT $NO_2$ mixing ratio (17). Several studies have applied this cloud-slicing method to infer FT $NO_2$ from low Earth orbit (LEO) satellite instruments, including OMI (17-19) and more recently TROPOMI (11, 20). However, the sparse sampling from LEO limits the ability to collect adjacent fully cloudy scenes. Marais*, et al.* (19) found that OMI-based cloud-sliced FT $NO_2$ products required averaging over thousands of kilometers to achieve significant correlations with aircraft $NO_2$ observations. Horner*, et al.* (11) developed a seasonal 1°×1° resolution product from TROPOMI, but its evaluation against aircraft observations was limited to broad regional averages.

Here we show that $NO_2$ column observations from the new TEMPO (Tropospheric Emissions: Monitoring of Pollution) geostationary satellite instrument launched in April 2023 provide an unprecedented capability to measure FT $NO_2$ by cloud slicing with high observation density and accuracy. TEMPO is the first geostationary satellite instrument to monitor atmospheric composition over North America (21). It provides unique hourly daytime $NO_2$ and cloud observations with finer spatial resolution (2.0×4.75 $km^2$ at 33.5°N) than LEO instruments, enabling a first observation of the diurnal cycle of FT $NO_2$ to test our understanding of $NO_x$ sources and chemistry. Combining the TEMPO cloud-sliced FT $NO_2$ observations with concurrent lightning observations from the Geostationary Lightning Mapper (GLM) (22) enables evaluation of the lightning $NO_x$ source parameterizations in current atmospheric chemistry models. Aside from its intrinsic interest, an accurate cloud-sliced FT $NO_2$ product is also critical for using the satellite measurements of $NO_2$ columns to infer surface $NO_x$ emissions in North America, because the FT makes an important and sometimes dominant contribution to the tropospheric $NO_2$ column that needs to be separated from the PBL (23-26).

**Results and Discussion**

**Free tropospheric $NO_2$ observations from TEMPO**

We define the FT as the 700-300 hPa pressure range and derive FT $NO_2$ mixing ratios on a 0.5°×0.625° (~50×50 $km^2$) grid from the regression slope of above-cloud $NO_2$ columns versus cloud pressures for TEMPO fully cloudy pixels. Figure S1 illustrates this FT $NO_2$ derivation for a 50×50 $km^2$ domain sampled by TEMPO over Boston (December 18, 2023, at 11:15 am), illustrating the high density of available information. We perform such retrievals for all suitably cloudy scenes in the full-year TEMPO record from 1 October 2023 to 30 September 2024 and aggregate the results to produce a seasonal 2°×2.5° product (*Materials and Methods*).

Figure 1 (left column) shows the spatial distribution and seasonal mean variability of FT $NO_2$ over North America observed by TEMPO. The average FT $NO_2$ varies from 13 ppt in winter to 24 ppt in spring-summer, in contrast to surface $NO_2$ which peaks in winter when the $NO_x$ lifetime is longest (27). The spring-summer maximum in FT $NO_2$ is consistent with lightning providing the dominant source, as found in previous model studies (6, 28, 29). FT $NO_2$ also shows a smoother spatial variability than surface $NO_2$ that we attribute to a longer lifetime and chemical recycling. This smoother spatial variability is also seen in aircraft observations (23). The highest FT $NO_2$ concentrations are found over regions with frequent lightning (eastern US during spring-summer), strong surface pollution uplift (Mexico City year-round), and intense fire events (southern Mexico in spring; Fig. S2). Wintertime data in Figure 1 are limited due to fewer high-reaching clouds. The FT $NO_2$ derived from cloud slicing corresponds to fully-cloudy conditions, but model results find no significant difference in $NO_2$ concentrations with all-sky conditions (20).

Figure 2A evaluates our seasonal TEMPO FT $NO_2$ product with daytime observations from aircraft campaigns averaged over the 2°×2.5° grid. These include older campaigns conducted in 2006-2015 (INTEX-B, ARCTAS, DC-3, $SEAC^4RS$, DISCOVER-AQ, FRAPPE, WINTER) and the recent AEROMMA in 2023 (Table S1). The aircraft observations are of NO concentrations, with NO$_2$ inferred from NO/NO$_2$ photochemical steady state (PSS), to avoid positive artifacts from thermally unstable nitrates in the NO$_2$ measurement (6). We removed high-NO$_2$ plumes from fires, lightning, and convected pollution outflow that would be preferentially targeted by the aircraft (*Materials and Methods*). We find an overall good spatial/seasonal correlation between TEMPO and the aircraft observations, considering the mismatches in observing strategy and years (r = 0.61, regression slope = 0.89, NMB = 14%). The variability in the aircraft observations is largely driven by higher values in spring and summer over the eastern US, which is captured by TEMPO and attributable to lightning.

We use the difference statistics between TEMPO and aircraft observations in Figure 2A to estimate an accuracy of 14% and precision of 50% for our FT NO$_2$ product. This precision estimate is conservative because it assumes that all of the difference between our FT NO$_2$ product and the aircraft observations is due to error in the FT NO$_2$ product. Errors and mismatch in the aircraft observations would also contribute. See *Materials and Methods* for further discussion of errors.

Our TEMPO FT NO$_2$ product vastly outperforms the previous LEO OMI cloud-sliced product, which exhibited poor correlation with aircraft observations at 8°×10° resolution (17) and required continental-scale averaging to achieve satisfactory agreement (19). Horner*, et al.* (11) more recently reported a global FT NO$_2$ cloud-sliced product at 1°×1° resolution using TROPOMI, which is also in LEO but has higher spatial resolution (3.5×5.5 km$^2$) than OMI (13×24 km$^2$). They evaluated that product at broad regional scales. We show in Fig. 2B that it also achieves significant correlation with the aircraft observations over North America on the 2°×2.5° grid (r = 0.41, regression slope = 0.73, NMB = -4%) though not as good as TEMPO. TROPOMI values average 20% lower than TEMPO, which we attribute in part to a 13:30 overpass local time (LT) and in part to a geometric air mass factor (AMF) that ignores atmospheric scattering in the NO$_2$ retrieval (*Materials and Methods*). Comparison of spatial and seasonal variability between the TEMPO and TROPOMI products on the 2°×2.5° grid shows a high degree of consistency (Fig. S3 and Table S3) but TEMPO has a smoother background and better coverage in the western US and in winter, even with only 1 year of data as compared to 4 years of data for TROPOMI.

**Model errors in lightning NO$_x$ emissions revealed by TEMPO and GLM**

Figure 1 compares TEMPO FT NO$_2$ observations with NASA Goddard Earth Observation System – Composition Forecast (GEOS-CF) model results, sampled by averaging the NO$_2$ mixing ratios over the 700-300 hPa column for the same scenes. GEOS-CF uses the GEOS-Chem chemical module (emissions, chemistry, deposition) within the GEOS meteorological model and data assimilation system (30). It provides NO$_2$ vertical shape factors for the AMFs in the TEMPO retrievals. The comparison in Figure 1 shows general consistency in magnitudes, spatial distributions, and seasonal variations, but there are important differences. GEOS-CF averages 5-10 ppt higher, mainly due to hotspots that are not present in the TEMPO observations. These hotpots tend to be associated with urban and/or high-altitude terrain where 700 hPa might still be within the model PBL, whereas the TEMPO cloud-sliced product would exclude the PBL by its requirement of full cloud cover. In addition, the GEOS-CF simulation uses the Year 2010 HTAP v2.2 inventory for anthropogenic NO$_x$ emissions (31), which would overestimate Year 2023 emissions in the US by a factor of 2.3 (32). GEOS-CF exaggerates FT NO$_2$ over regions of western North America with large 2024 fires (Fig. S2). It assumes that 35% of fire emissions are directly injected into the FT (3.5-5.5 km), which is likely excessive (33, 34). In addition, GEOS-CF and other models do not properly account for the observed rapid conversion of NO$_x$ to organic nitrate species in fire plumes (35).

GEOS-CF overestimates FT NO$_2$ over land in the eastern US while underestimating FT NO$_2$ offshore during spring and summer, suggesting large errors in the model's parameterization of lightning emissions. This is supported by GLM observations from the NOAA Geostationary

Operational Environmental Satellites (GOES) (*Materials and Methods*), as shown in Figure 3A. GLM detects a high frequency of lightning flashes offshore from the eastern US, with a spatial pattern similar to TEMPO FT $NO_2$. In GEOS-CF, lightning flashes are parameterized using a standard model scheme that takes cloud top height (CTH) as a proxy, with different coefficients assigned for land and ocean (Figure 3B) (36). This differentiation is intended to account for higher updraft velocities over land, but we find that it underestimates lightning flash frequency and resulting FT $NO_2$ levels over the North Atlantic. The GLM observations show a factor of 10 diurnal variation in lightning flash frequency over land from the minimum at 8 LT to the maximum at 16 LT, but GEOS-CF has little diurnal variation. GEOS-CF further applies a $NO_x$ yield of 500 mol N flash$^{-1}$ for regions north of 35°N, doubling the value of 260 mol N flash$^{-1}$ used elsewhere (Fig. S4), which likely contributes to excessive FT $NO_2$ over land. This adjustment was originally implemented in GEOS-Chem to match FT $NO_2$ observations from an older aircraft campaign (ICARTT) (37), but that dataset may have been biased high due to positive interference from nitrate reservoirs (6). Overall, we see that the combination of TEMPO and GLM observations offers a powerful resource for improving lightning $NO_x$ emission parameterizations in models.

**First diurnal observations of FT $NO_2$**

TEMPO enables the first-ever observations of the diurnal variation of FT $NO_2$, as shown in Figure 4. The data show little diurnal variation in winter and a pronounced midday minimum in other seasons. This FT $NO_2$ diurnal cycle differs from that previously reported from geostationary observations for the urban PBL, where the combination of emissions and chemistry drives a $NO_2$ increase over the course of the day in winter and a weak afternoon minimum in summer (38, 39). GEOS-CF reproduces well the diurnal variation of FT $NO_2$ observed by TEMPO and its seasonality except that it exaggerates the morning decrease in summer and fall. The midday minimum outside of winter can be explained by photochemical oxidation by the OH radical and a diurnal variation in the $NO/NO_2$ steady-state ratio. The morning overestimate in GEOS-CF may be caused by excessive lightning at night and in early morning, as shown by comparison to the GLM diurnal cycle where observed lightning during these hours is particularly low in summer-fall (Fig. 3C-D). More work is needed to understand the interplay between sources and chemistry determining FT $NO_2$ distributions, but we see that the diurnal variation observed by TEMPO provides unique information. The availability of hourly FT $NO_2$ profiles also supports more accurate quantification of PBL $NO_x$ emission diurnal patterns by enabling subtraction of the FT contribution from total tropospheric $NO_2$ columns.

In summary, we have presented a new cloud-sliced FT $NO_2$ product from the TEMPO geostationary satellite instrument with high quality, high resolution, and the first diurnal coverage over North America. Combined with independent lightning observations from GLM, our data confirm that lightning is the main source of FT $NO_2$ outside winter and reveal large errors in the standard lightning $NO_x$ emission schemes commonly used in atmospheric chemistry models. They provide a unique resource to improve the lightning $NO_x$ emission parameterization in models and from there our global understanding of tropospheric ozone and OH.

**Materials and Methods**

**TEMPO Satellite Observations.** TEMPO is a geostationary satellite instrument designed for monitoring of air quality over North America. It is on board the IntelSat-40e satellite launched in April 2023 and stationed at 91°W. It provides hourly daytime measurements over North America from 17°N to 60°N with a pixel resolution of 2.0×4.75 km$^2$ at 33.5°N. TEMPO measures solar backscatter in the UV/Vis wavelength range (293-494 nm and 538-741 nm), enabling the retrieval of a suite of trace gases, aerosols, and cloud parameters (21). Here we use the TEMPO L2 V03 cloud (40) and $NO_2$ (41) products from October 2023 (the beginning of nominal operations) to September 2024.

Cloud parameters are retrieved by TEMPO to support trace gas retrievals (42). We use two parameters: cloud radiance fraction (CRF), derived at 466 nm, and cloud optical centroid pressure (OCP), retrieved from the $O_2$-$O_2$ column near 477 nm. CRF is the fraction of observed radiance reflected by the cloud. OCP represents the pressure level where a proxy single-layer reflective cloud (with an albedo of 0.8) would be placed to quantify photon path length (43, 44). We define fully cloudy scenes as those with CRF>0.9.

TEMPO $NO_2$ slant column densities (SCDs) along the optical path are retrieved by spectral fitting of backscattered solar radiances in the 405-465 nm window (45). For fully cloudy scenes, the observed SCDs are for the column above the OCP. We then derive the above-cloud $NO_2$ vertical column densities (VCDs) by dividing the SCDs by the air mass factor AMF$_{cloudy}$ (18, 24), calculated as:

$$\text{AMF}_{cloudy} = \int_{OCP}^{TOA} w(p)S(p)dp, \quad [1]$$

where $w(p)$ is the pressure-dependent scattering weight obtained from a pre-computed lookup table characterizing the instrument sensitivity to $NO_2$ at different altitudes, and $S(p)$ represents the relative vertical profile of $NO_2$ (shape factor) provided by the GEOS-CF model. Both $w(p)$ and $S(p)$ are provided in the TEMPO L2 $NO_2$ files. The integration is performed from the OCP to the top of atmosphere (TOA). For our analysis, we use above-cloud $NO_2$ VCDs for fully cloudy scenes (CRF>0.9) with OCP between 700 and 300 hPa, excluding those with quality flags > 0, solar zenith angle > 70°, surface albedo > 0.3, or snow/ice fraction > 0.

**TEMPO cloud-slicing.** The cloud-slicing technique infers the FT $NO_2$ mixing ratio from the slope of the regression line for above-cloud $NO_2$ VCDs versus cloud OCPs in adjacent fully cloudy scenes, assuming horizontal homogeneity in the $NO_2$ vertical profiles (17, 18). We perform these regressions for each TEMPO scan over 0.5°×0.625° domains (50×50 km$^2$), selecting only those that contain at least 30 fully cloudy pixels, an OCP range exceeding 200 hPa and with sufficient variability (standard deviation > 67 hPa), and a uniform $NO_2$ stratospheric column (relative standard deviation < 0.02). We apply a Theil regression to minimize the influence of outliers (11). The regression slope is converted to $NO_2$ mixing ratio as described in Choi, et al. (17). The error standard deviation of the slope is obtained with bootstrap resampling. Slopes that are negative beyond two standard deviations are removed (<5% of total estimates). From inspection of the data, we attribute these significantly negative slopes to horizontal $NO_2$ inhomogeneity within the 50×50 km$^2$ domain, where $NO_2$ lightning plumes from high-altitude clouds enhance $NO_2$ columns over these pixels but not over adjacent lower-altitude cloud scenes.

The FT $NO_2$ mixing ratios for individual 0.5°×0.625° scenes are averaged over a 2°×2.5° grid, weighted by the error standard deviations of the slopes, and the resulting 2°×2.5° scenes are further averaged seasonally. Each 2°×2.5° grid cell requires at least 20 FT $NO_2$ estimates from at least 10 different days to compute a seasonal average. When deriving the FT $NO_2$ diurnal product (Fig. 4), we relax the criteria to 10 estimates per season for a given hour of day.

Errors in our FT $NO_2$ product arise from the cloud retrievals, the $NO_2$ column retrievals, and the regression which assumes a uniform $NO_2$ mixing ratio within the 700-300 hPa layer over a 50×50 km$^2$ domain. The CRF retrieved by TEMPO may have a high bias of ~0.05 (42) but we find that a stricter threshold for fully cloudy scenes (CRF>0.95) does not change results significantly. OCP retrievals have uncertainties below 20 hPa for CRF > 0.9 (42). The below-cloud $NO_2$ contribution is typically <1% of total SCDs for CRF > 0.9 and has little impact on the results when subtracted upfront. The sensitivity of AMF$_{cloudy}$ to the vertical $NO_2$ shape factor is less than 5% (24) due to the nearly uniform scattering weights above clouds. We found that using a geometric AMF instead of AMF$_{cloudy}$ introduces a low bias of ~2 ppt (10%) for seasonal FT $NO_2$. The $NO_2$ SCDs retrievals have a precision about 5×10$^{14}$ molec cm$^{-2}$ after normalization by geometric AMF (45).

NO$_2$ retrievals in the early morning and late afternoon may be biased (45) but this would have little impact on the cloud-sliced results as the biases would cancel out in the regression. The error standard deviation on the regression slope for NO$_2$ VCD versus cloud OCP determined from bootstrap resampling has a median value of 60% for individual estimates, but that error would decrease with averaging.

To estimate an overall error statistic for our seasonal FT NO$_2$ product on the 2°x2.5° grid, we use the normalized mean bias (NMB) and standard deviation of the differences between TEMPO and aircraft observations (Fig. 2A). From there we infer an accuracy of 14% and precision of 50%. This is conservative because it assumes that the aircraft data are accurate and representative of the TEMPO observations. A significant fraction of the difference could instead originate from representation error considering that the aircraft campaigns were conducted with only limited sampling of vertical profiles and for different years.

**Aircraft Observations.** We use aircraft observations from the INTEX-B, ARCTAS, DC-3, SEAC$^4$RS, DISCOVER-AQ, FRAPPE, WINTER, and AEROMMA campaigns over North America to evaluate our cloud-sliced FT NO$_2$ product. These campaigns operated a DC-8 (up to 12 km), GV (up to 12 km), or C-130 (up to 7 km) aircraft (Table S1a). To compare with our cloud-sliced product, we focus on flight segments between 3-9 km altitude (~700-300 hPa) with a minimum vertical profiling of 3 km altitude within each 2°×2.5° grid cell. The two winter campaigns, DISCOVER-AQ (California flights) and WINTER, have limited vertical extent (with ceilings of 3.2-5 km), and for those we include all data above 3 km without applying the vertical extent filter.

Table S1b lists the measurements used in this work. NO$_2$ measurements in the free troposphere can be affected by positive interference from the thermal decomposition of labile NO$_x$ reservoirs due to higher instrument temperatures compared to ambient air (46-49). To avoid these artifacts and following approaches from previous studies (6, 11, 50), we calculate photochemical steady state (PSS) NO$_2$ from NO measurements by assuming a dynamic daytime equilibrium between fast oxidation of NO and photolysis of NO$_2$. NO oxidation in the FT is mainly driven by O$_3$, with a smaller contribution from HO$_2$ (14, 50). PSS NO$_2$ is then calculated as:

$$[NO_2]_{PSS} = [NO] \times \frac{k_1[O_3] + k_2[HO_2]}{j_{NO_2}}, \quad [2]$$

where $k$ represents rate constants (51), [] denotes number density, and $j_{NO2}$ is the NO$_2$ photolysis frequency. PSS NO$_2$ is computed from aircraft measurements, except for [HO$_2$] in campaigns where it was not measured (Table S1b). In these cases, $k_2$[HO$_2$] is estimated based on its relative contribution to ($k_1$[O$_3$] + $k_2$[HO$_2$]) using data from other campaigns where [HO$_2$] measurements are available. For the AEROMMA campaign, we use directly measured NO$_2$ from a newly developed laser-induced fluorescence (LIF) instrument (52), in which the NO$_2$ photolytic converter is maintained at near-ambient temperatures to minimize interference.

We use only daytime aircraft observations (9-16 LT for winter and 8-17 LT for other seasons) for consistency with the TEMPO observing time window. Additional filtering is applied to exclude observations affected by fresh convection (number concentration of condensation nuclei larger than 10 nm > 5×10$^3$ cm$^{-3}$), fresh NO$_x$ emissions (NO$_y$/NO < 3 mol mol$^{-1}$), and biomass burning plumes (CO > 200 ppbv and CH$_3$CN > 200 pptv; CO > 500 ppbv or CH$_3$CN > 500 pptv if one of these observations is unavailable). Flights focused on sampling deep convection or fire plumes are also excluded (Table S1a) (53-55). We then average the FT NO$_2$ concentrations within each 2°×2.5° grid cell for each flight and compute the seasonal mean for each grid cell across all flights.

**GOES GLM Observations.** We use satellite-based lightning flash observations from the two Geostationary Lightning Mapper (GLM) instruments onboard GOES-16 and GOES-18, with GOES-16 covering the eastern Americas (centered at 75.2°W), GOES-18 covering the western Americas (centered at 137.2°W), and both providing latitudinal coverage from 54°S to 54°N. GLM is an optical imager that detects cloud top lightning illumination at 777.4 nm every 2 ms, with a nadir pixel resolution of 8 km (22). The GLM retrieval algorithm defines a lightning event as the occurrence of a single pixel exceeding the background radiance threshold. It further aggregates simultaneous adjacent events into groups and clusters sequential groups separated by less than 330 ms and 16.5 km into a single flash (56). The GLM Level 2 product reports the radiance-weighted centroids of lightning flash locations, which we map to the 0.5°×0.625° grid. To minimize the degradation in data quality near the edge of each instrument's field of view, we use GLM data from GOES-16 for regions east of 106.2°W and from GOES-18 for regions west of 106.2°W (57).

**Data, Materials, and Software Availability.** The TEMPO cloud and $NO_2$ products are publicly available at https://asdc.larc.nasa.gov/project/TEMPO?level=2. Aircraft observations for the INTEX-B, ARCTAS, DC-3, SEAC$^4$RS, DISCOVER-AQ, and FRAPPE campaigns are available at https://www-air.larc.nasa.gov/missions/merges/. Data for the WINTER campaign are available at https://www.eol.ucar.edu/field_projects/winter, and data for the AEROMMA campaign are available at https://csl.noaa.gov/projects/aeromma/data.html. GLM lightning flash L2 data are available at https://noaa-goes16.s3.amazonaws.com/index.html#GLM-L2-LCFA/. GEOS-CF simulated $NO_2$ vertical profiles are included with the TEMPO $NO_2$ product. Emission input data for GEOS-CF are publicly available at https://portal.nccs.nasa.gov/datashare/gmao/geos-cf/v1/. TROPOMI FT $NO_2$ is on the UCL Data Repository (https://doi.org/10.5522/04/28846736). TEMPO FT $NO_2$ seasonal product is archived at Zenodo (https://zenodo.org/records/15299339) and is currently under embargo. The dataset will be made publicly available upon publication.

**Acknowledgments**

We are grateful to the instrument teams of the INTEX-B, ARCTAS, DC-3, SEAC$^4$RS, DISCOVER-AQ, FRAPPE, WINTER, and AEROMMA campaigns for making their data freely available. This work was supported by the Harvard-NUIST Joint Laboratory for Air Quality and Climate. TEMPO operations and data processing at the Smithsonian Astrophysical Observatory are supported by NASA contracts NNL13AA09C and 80MSFC24CA004. EAM acknowledges support from the European Research Council under the European Union's Horizon 2020 research and innovation programme (through a Starting Grant awarded to Eloise A. Marais, UpTrop [grant no. 851854])

**Figures and Tables**

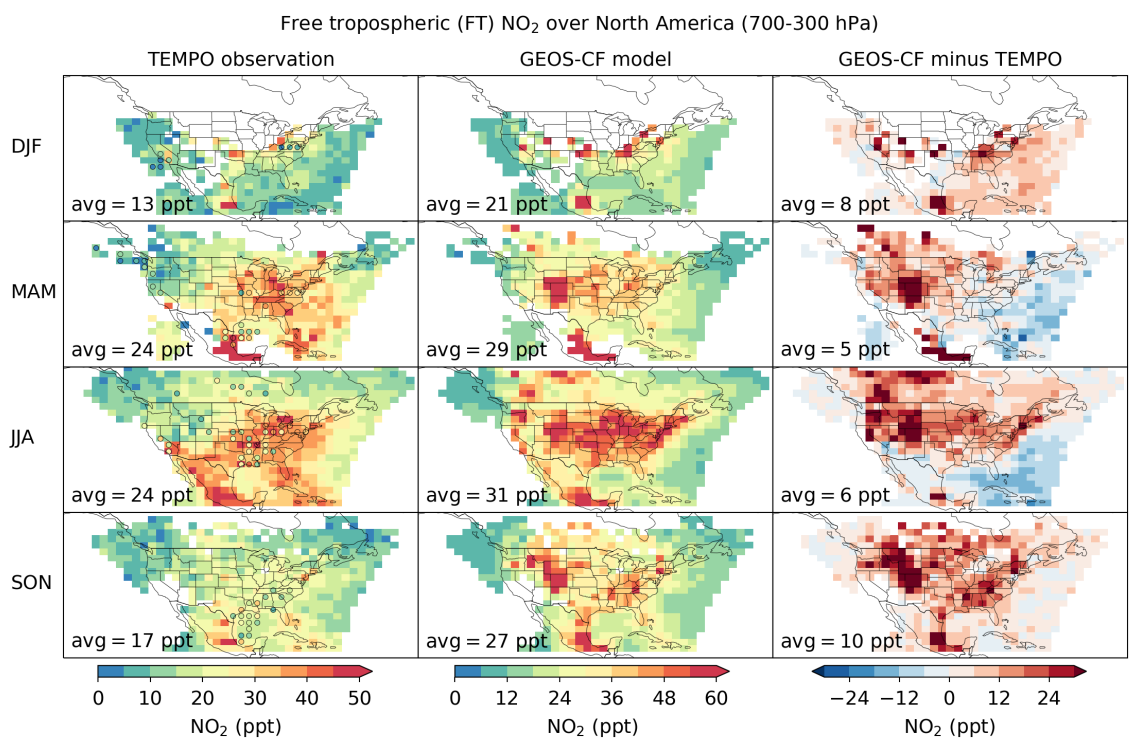

**Figure 1.** Free tropospheric (FT, 700-300 hPa) NO$_2$ mixing ratio over North America for different seasons of October 2023 – September 2024. TEMPO observations averaged seasonally on a 2°×2.5° grid are compared to observations from a collection of aircraft campaigns in different years (circles) and to model results from Goddard Earth Observation System – Composition Forecasts (GEOS-CF) sampled at the same locations and times. The difference between the two is shown in the third column. White areas are either outside the TEMPO field of regard (FOR) or do not have sufficient full-cloud scenes for application of the cloud-sliced method. Spatial average values for the observation domains are shown inset. The aircraft campaigns are listed in Table S1 and further evaluation of the TEMPO product with these observations is in Figure 2.

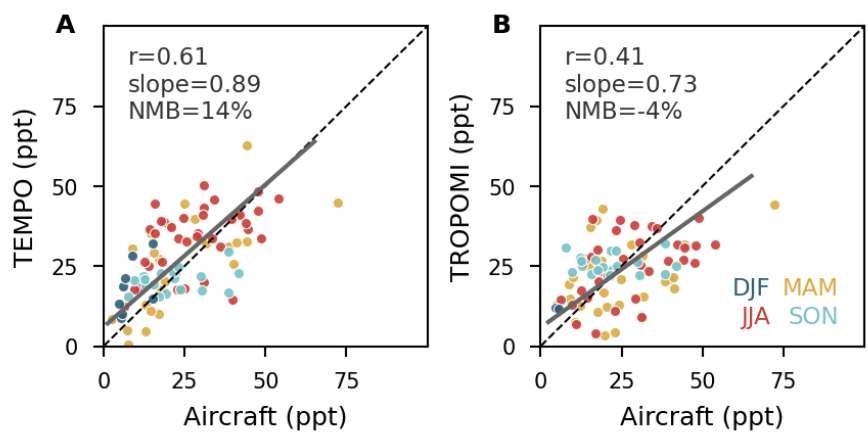

**Figure 2.** Comparison of cloud-sliced FT NO$_2$ with aircraft observations over North America. (A) TEMPO product as described in Figure 1. (B) TROPOMI product from Horner, *et al.* (11) for June 2018–May 2022, processed at the same pressure range and spatial resolution as TEMPO. Aircraft observations are from the collection of INTEX-B, ARCTAS, DC-3, SEAC$^4$RS, DISCOVER-AQ, FRAPPE, WINTER, and AEROMMA campaigns, averaged seasonally on a 2°×2.5° grid. Comparison statistics include the correlation coefficient (r), reduced-major-axis (RMA) regression line and slope, and normalized mean bias (NMB). The 1:1 line is shown as dashed.

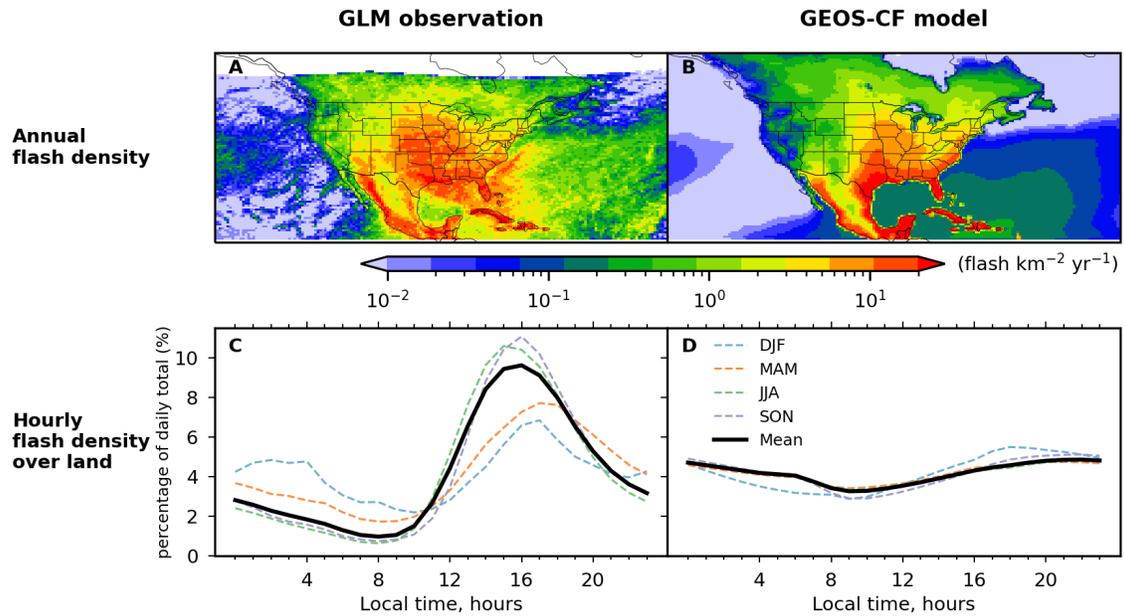

**Figure 3.** Lightning flash densities over North America observed by the Geostationary Lightning Mapper (GLM, left column) and simulated by the GEOS-CF model (right column). The top panels show the annual flash density distribution, and the bottom panels show the relative diurnal variations of seasonal and annual flash density over land. GLM observations are from the GOES-16 and GOES-18 Level 2 product (*Materials and Methods*). No GLM observations are available north of 54°N (in white). Results cover the period from October 2023 to September 2024.

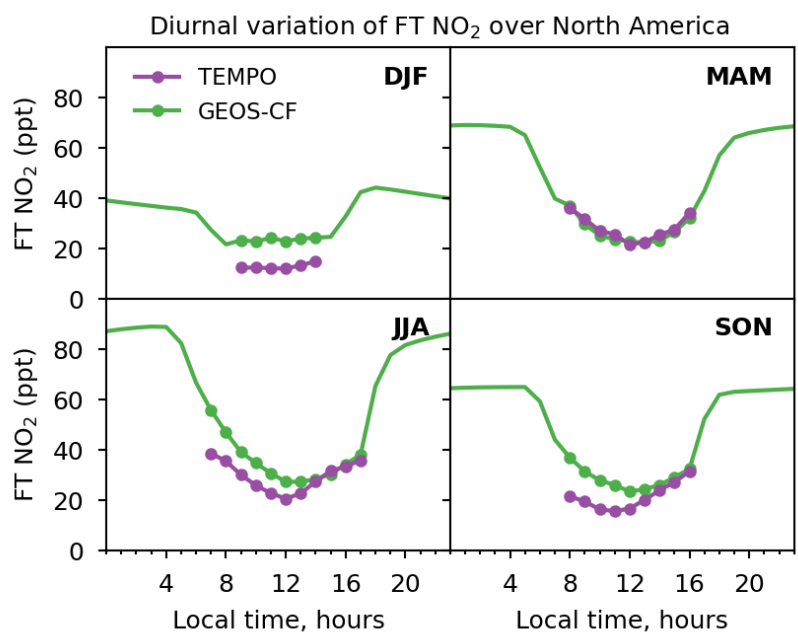

**Figure 4.** Diurnal variation of free tropospheric (FT, 700-300 hPa) NO$_2$ mixing ratios averaged over the TEMPO domain for different seasons. TEMPO cloud-sliced observations are available for the hours indicated by symbols (9-14 LT for winter; 8-16 LT for spring and fall; 7-17 LT for summer) and are compared with GEOS-CF simulations sampled at the same locations and times. Outside of the TEMPO observing window, GEOS-CF results are shown as seasonal averages under all-sky conditions. The averages over the TEMPO domain are computed from the mean mixing ratios in 2°×2.5° grid cells, and only grid cells with at least 10 seasonal observations for each hour are considered in the analysis.